\documentstyle[12pt]{article}

\newcommand \kakko[1]{\left[ {#1} \right]}
\newcommand \ckakko[1]{\left\{ {#1} \right\}}
\newcommand \utilde[1]{\raisebox{-.8em}{$\tilde{~}$} \!\!\! {#1}}

\newcommand \ignore[1]{}

\newcommand \fra[2]{\displaystyle
{\frac{\textstyle {#1}}{\textstyle {#2}}}}

\makeatletter
\@addtoreset{equation}{section}
\makeatother

\begin{document}

\begin{flushright}
  UT-622 \\
  November 1992 \\
\end{flushright}
\vspace{24pt}
\begin{center}
\begin{large}
{\bf
The Non-perturbative Canonical Quantization of the N=1 Supergravity
}
\end{large}

\vspace{36pt}
        Takashi Sano\footnote{
        sano@tkyvax.phys.s.u-tokyo.ac.jp}
        and Jun' ichi Shiraishi\footnote{
        shiraishi@tkyvax.phys.s.u-tokyo.ac.jp}

\vspace{6pt}

{\it Department of physics, University of Tokyo }\\
{\it Bunkyo-ku, Tokyo 113, Japan }

\vspace{48pt}

\underline{ABSTRACT}

\end{center}

\vspace{4cm}

The non-perturbative canonical quantization
of the N=1 supergravity with the non-zero cosmological
constant is studied using the Ashtekar formalism.
A semi-classical wave function is obtained
and it has the form of the exponential
of the N=1 supersymmetric extension of the Chern-Simons
functional. The N=1 supergravity in the Robertson-Walker universe
is also examined and some analytic solutions are obtained.

\vfill
\newpage

\ignore{}

\section{Introduction}

 Recently several attempts have been made at constructing
quantum theory of gravity and proved successful in the case of low dimensional
gravity theories.
In particular, a major progress has been achieved in
the 2-dimensional quantum gravity based on the methods of
the conformal field theory. In the 3-dimensional case, we also
have various approaches to the quantum gravity such as
the Chern-Simons gauge theory and the Turaev-Viro theory, and so on.
The 4-dimensional quantum gravity has, however, proved too difficult so far to
be constructed completely. The main difficulties are that
it is unrenormalizable and highly non-linear.
While the 3-dim gravity is also unrenormalizable,
it has no dynamical degrees of freedom and we can formulate it
as a topological field theory. On the other hand
the 4-dim gravity has the gravitons and
can't be described by some topological field theory straightforwardly.

There are several approaches to overcome these difficulties of
the 4-dim quantum gravity. One of them is to find the renormalizable
theory, which contains the Einstein gravity in a suitable limit.
The typical one of this approach is the superstring theory.
The stand point of this approach is to modify the Einstein
gravity such that it has more tractable perturbative behaviour.
On the other hand, we can take the position that the
quantum gravity should be defined non-perturbatively.
There are two well-known attempts to construct
the non-perturbative quantum gravity.
One is the lattice gravity and the other is the canonical quantization
by the ADM formalism \cite{ADM}.

The lattice gravity has been successful in 2 or 3 dimension by
means of the random triangulation method. The 4-dimensional lattice
gravity is now in progress. But it has not been ascertained in 4 dimension
whether we
can take a continuum limit.
The ADM canonical formalism has
its own difficulty.
As is well-known, the Hamiltonian of the Einstein-Hilbert action
reduces to the constraints in the ADM formalism. These constraints
are the complicated non-polynomials of the canonical variables.
So the canonical quantization can be carried out only
in the cases in which there are few degrees of freedom.

Recently Ashtekar has presented a new formulation
of the Einstein gravity \cite{AA}\cite{AAA}. In this formalism,
all the constraints of
the gravity are simple polynomials of the canonical variables, so
we would expect to solve the non-perturbative quantum gravity.
In fact some wave functions and the physical states of the gravity are derived
\cite{JS}\cite{HK}.
The Ashtekar formalism can be extended to the N=1,2 supergravities and
the constraints are again polynomials of the canonical variables
\cite{J}\cite{KS}.

The aim of this paper is to consider the non-perturbative
canonical quantization of the N=1 supergravity by the Ashtekar formalism.
While the ADM formalism of the N=1 supergravity has been investigated
in Ref.\cite{D}, it is extremely more complicated than that of the ordinary
gravity, so it seems unsuitable for quantization.
As we will see soon,
the Ashtekar formalism gives us a powerful aid for our aim.
Especially we give attention to the case that the theory has the
non-zero cosmological constant. The case of the Einstein gravity in the
same condition have been considered by Kodama\cite{HK}.

This paper is organized as follows. In section 2, we give
the brief review of the Ashtekar formalism of the N=1 supergravity.
After the 3+1 decomposition of the action, we get the constraints of
the supergravity. We solve these constraints semi-classically and
obtain the {\it{holomorphic wave function}} of the N=1 supergravity.
We consider the special case that the metric is
given by the Robertson-Walker metric in section 3.
We solve the Wheeler-Dewitt
equation and again obtain the semi-classical wave function of the universe.
In section 4, we derive the equations which determine the
classical limit of the quantum Robertson-Walker universe
and obtain the several analytic solutions. We examine
these equations by the numerical simulation.
Section 5 is devoted to the discussion.
The notations and the formulas used in this paper are given in the appendix.

\section{The Ashtekar Formalism and the WKB wave function
of N=1 Supergravity}

In this section we present the Ashtekar formalism of the N=1 supergravity
and solve the constraints. The Ashtekar formalism of the N=1 supergravity
has been given first by Jacobson\cite{J} and reformulated in more
elegant form in ref\cite{CDJM}.

{}From now on we use the method of\, the 2-form gravity\cite{CDJM}. We
represent
the left- and the right-spinor indices as $A, B, C,\cdots$ and
$A', B', C',\cdots$, respectively. $e^{AA'}, \, \psi_{A},$ and
$ \psi_{A'}$ express the vierbein, the left- and the right-component
of the gravitino, respectively. We define the 2-form fields
$\Sigma^{AB}$ and $ \chi^{A}$ as
\begin{eqnarray}
            \Sigma^{AB} &=& e^{A}_{A'}  \wedge e^{A'B}    \label{eq:Sigma}, \\
            \chi^{A}    &=& e^{A}_{A'}  \wedge \psi^{A'}. \label{eq:chi}
\end{eqnarray}
Now the chiral Lagrangian of the N=1 supergravity is given as \cite{KS}:
\begin{eqnarray}
       -i{\it{L}} &=& -i{\it L_{0}} -i{\it L_{cosm}},\label{eq:L} \\
       -i{\it L_{0}} &=&\Sigma_{AB}\wedge R^{AB}
                      +\chi^{A}\wedge D\psi_{A}
                      -\frac{1}{2}\Psi_{ABCD}\Sigma^{AB}\wedge\Sigma^{CD}
                      -\kappa_{ABC}\Sigma^{AB}\wedge\chi^{C},
                  \nonumber \\
       -i{\it L}_{cosm} &=&  - \frac{g^{2}}{6} \Sigma_{AB} \wedge \Sigma^{AB}
                      + \frac{1}{2} \lambda g \, \Sigma^{AB}
                      \wedge \psi_{A} \wedge \psi_{B}
                      -\frac{g}{6 \lambda} \chi_{A}
                      \wedge \chi^{A},\nonumber
\end{eqnarray}
where
$R_{AB}$ is the curvature of the anti-self-dual part of $ SO(3,1) $
connection $\omega_{AB}$, $ D $ is the covariant derivative
with respect to $\omega_{AB}$, and $ g $ and $\lambda $ are the real constants.
When we add cosmological term $-\frac{g^{2}}{6}\Sigma_{AB}\wedge\Sigma^{AB}$
to Lagrangian, we must add other terms appearing in ${\it L_{cosm}}$\cite{PK}.
$\lambda$ can be regarded as the gravitational constant.
Cosmological constant is given by $\Lambda=g^{2}$.
$\Psi_{ABCD}$ and $\kappa_{ABC}$ are the Lagrange multipliers by which we
require the algebraic constraints
\begin{eqnarray}
                \Sigma^{(AB} \wedge \Sigma^{CD)} &=& 0, \label{eq:cons1}\\
                \Sigma^{(AB} \wedge \chi^{C)} &=& 0, \label{eq:cons2}
\end{eqnarray}
where the indices between `(' and `)' are completely
symmetrized, and these equations
guarantee the decomposition (\ref{eq:Sigma}) and (\ref{eq:chi}).

In this paper we consider N=1 supergravity with non-zero cosmological term.

The Lagrangian (\ref{eq:L}) has the left- and the right local supersymmetries.
The left
supersymmetry transformation is given by:
\begin{eqnarray}
                \delta_{L}\Sigma^{AB} &=&
                -\chi^{(A}\epsilon^{B)}, \nonumber \\
                \delta_{L}\omega_{AB} &=&
                \lambda g \psi_{(A}\epsilon_{B)}, \nonumber \\
                \delta_{L}\psi_{A} &=&
                D \epsilon_{A}, \nonumber \\
                \delta_{L}\chi^{A} &=&
                \lambda g \Sigma^{AB}\epsilon_{B}, \nonumber \\
                \delta_{L}\kappa_{ABC} &=&
                -\Psi_{ABCD}\epsilon^{D}, \nonumber \\
                \delta_{L}\Psi_{ABCD} &=&
                -2\lambda g \,\kappa_{(ABC}\epsilon_{D)},
\end{eqnarray}
where $\epsilon_{A}$ is the fermionic 0-form parameter.
The right transformation has the peculiar form:
\begin{eqnarray}
                \delta_{R}\Sigma^{AB} &=&
                \psi^{(A}\wedge \eta^{B)}, \nonumber \\
                \delta_{R}\omega_{AB} &=&
                -\kappa_{ABC}\eta^{C}, \nonumber \\
                \delta_{R}\psi_{A} &=&
                \frac{g}{3\lambda}\eta_{A}, \nonumber \\
                \delta_{R}\chi_{A} &=&
                -D\eta_{A},\nonumber \\
                \delta_{R}\kappa_{ABC} &=& 0, \nonumber \\
                \delta_{R}\Psi_{ABCD} &=& 0,
\end{eqnarray}
where the parameter $\eta_{A}$ is the fermionic 1-form parameter which
satisfies the algebraic constraint
\begin{equation}
                \Sigma^{(AB} \wedge \eta^{C)}=0,
\end{equation}
which can be solved on shell as
\begin{equation}
                \eta^{A}\sim e^{A}_{A'}\epsilon^{A'}.
\end{equation}
Now we rewrite the Lagrangian (\ref{eq:L}) in the canonical form.
First we define the variables $\tilde{\pi}^{iAB}$ and
$\tilde{\pi}^{iA}$ as
\begin{eqnarray}
                \tilde{\pi}^{iAB} &=&
                \frac{1}{2}\epsilon^{ijk}
                \Sigma^{AB}_{jk}, \\
                \tilde{\pi}^{iA} &=&
                \frac{1}{2}\epsilon^{ijk}\chi^{A}_{jk}.
\end{eqnarray}
Then the algebraic constraints (\ref{eq:cons1}), (\ref{eq:cons2})
 can be written as
\begin{eqnarray}
              &&  \Sigma^{(AB}_{0i}\tilde{\pi}^{CD)i} = 0, \\
              &&  \Sigma^{(AB}_{0i}\tilde{\pi}^{C)i}
                +\tilde{\pi}^{i(AB}\chi^{C)}_{0i} = 0. \label{eq:App1}
\end{eqnarray}
As is shown in the appendix, these equations can be solved as
\begin{eqnarray}
                \Sigma^{AB}_{0i} &=& -\frac{1}{2}
                \epsilon_{ijk} \kakko{-i \;\utilde{N}
                \tilde{\pi}^{jA}_{C} \tilde{\pi}^{kCB}
                +2N^{j} \tilde{\pi}^{kAB} },  \label{eq:App2} \\
                \chi^{A}_{0i} &=& - \epsilon_{ijk} \kakko{
                -i \;\utilde{N} \tilde{\pi}^{jA}_{B} \tilde{\pi}^{kB}
                +N^{j}\tilde{\pi}^{kA}}
                +\epsilon_{ijk}\tilde{\pi}^{jA}_{B}
                \tilde{\pi}^{kBC}\utilde{M}_{C}, \label{eq:App3}
\end{eqnarray}
where $\utilde{M}_{A}$ is the fermionic field of the weight -1, and
$\utilde{N}$ and $N^{i}$ correspond to the lapse function
and the shift vector in the ADM formalism, respectively.

The Lagrangian rewritten in the canonical form is
\begin{eqnarray}
                -iL &=& \tilde{\pi}^{iAB}\dot{\omega}_{iAB}
                +\tilde{\pi}^{iA}\dot{\psi}_{iA} \nonumber \\
                &&+\omega_{0AB}{\bf G}^{AB}-\psi_{0A}{\bf L}^{A}
                +\utilde{M}_{A}{\bf R}^{A}+\frac{1}{2}\,i\;\utilde{N}{\bf H}
                -N^{i}{\bf H}_{i}. \label{eq:gl}
\end{eqnarray}
The coefficients $\omega_{0AB}$, $\psi_{0A}$, $\utilde{M}_{A}$,
$\utilde{N}$, and $N^{i}$ are the Lagrange multipliers and
the constraints are given by
\begin{eqnarray}
                {\bf G}^{AB} &=& D_{i}\tilde{\pi}^{iAB}
                -\psi^{(A}_{i}\tilde{\pi}^{B)i}, \label{eq:G} \\
                {\bf L}^{A} &=& D_{i}\tilde{\pi}^{iA}
                -\lambda g \tilde{\pi}^{iAB}\psi_{Bi}, \\
                {\bf R}^{A} &=& \tilde{\pi}^{iA}_{C}
                \tilde{\pi}^{jCB}
                \kakko{(D\psi_{B})_{ij}+\frac{g}{3\lambda}
                \epsilon_{ijk}\tilde{\pi}^{k}_{B}}, \\
                {\bf H} &=& \tilde{\pi}^{iA}_{C}\tilde{\pi}^{jCB}
                \kakko{R_{ijAB}-\frac{g^{2}}{3}\epsilon_{ijk}
                \tilde{\pi}^{k}_{AB}+\lambda g
                \psi_{i(A}\psi_{B)j}} \nonumber \\
                &&+2\tilde{\pi}^{iA}_{B}\tilde{\pi}^{jB}
                \kakko{(D\psi_{A})_{ij}+\frac{g}{3\lambda}
                \epsilon_{ijk}\tilde{\pi}^{k}_{A}},\label{eq:H} \\
                {\bf H}_{i} &=& \tilde{\pi}^{jAB}
                \kakko{R_{ijAB}-\frac{g^{2}}{3}\epsilon_{ijk}
                \tilde{\pi}^{k}_{AB}+\lambda g
                \psi_{i(A}\psi_{B)j}} \nonumber \\
                &&+\tilde{\pi}^{jA}\kakko{(D \psi_{A})_{ij}
                +\frac{g}{3\lambda}
                \epsilon_{ijk}\tilde{\pi}^{k}_{A}}. \label{eq:Hi}
\end{eqnarray}
The Poisson brackets between the canonical variables
\footnote[1]{We use the left derivatives for the fermionic fields.} are
\begin{eqnarray}
                \ckakko{\omega_{iAB}(x,t), \,\tilde{\pi}^{jCD}(y,t)}
                &=& -i\delta^{j}_{i}\delta^{C}_{(A}\delta^{D}_{B)}
                \delta^{(3)}(x-y), \label{eq:PB1} \\
                \ckakko{\psi_{iA}(x,t), \,\tilde{\pi}^{jB}(y,t)}
                &=& -i\delta^{j}_{i}\delta^{B}_{A}
                \delta^{(3)}(x-y). \label{eq:PB2}
\end{eqnarray}
${\bf G}^{AB}$, ${\bf L}^{A}$, ${\bf R}^{A}$, ${\bf H}$, and ${\bf H}_{i}$ are
the generators of
the local Lorentz transformation,
the left- and the right supersymmetry transformations,
the time evolution, and the 3-dim diffeomorphism, respectively.
These constraints are written in the polynomials of the canonical variables and
form the closed Poisson algebra under (\ref{eq:PB1})and (\ref{eq:PB2})\cite{J}.

Now we start the quantization. The (anti-)commutation relations are
\begin{eqnarray}
                \kakko{\omega_{iAB}(x,t),\,\tilde{\pi}^{jCD}(y,t)}
                &=& \delta^{j}_{i}\delta^{C}_{(A}\delta^{D}_{B)}
                \delta^{(3)}(x-y), \\
                \ckakko{\psi_{iA}(x,t),\,\tilde{\pi}^{jB}(y,t)}
                &=& \delta^{j}_{i}\delta^{B}_{A}\delta^{(3)}(x-y).
\end{eqnarray}
We choose the representation in which the variables $\omega_{iAB}$
and $\psi_{iA}$ are diagonalized.
\begin{eqnarray}
                \tilde{\pi}^{iAB} &=&
                -\frac{\delta}{\delta\omega_{iAB}}, \\
                \tilde{\pi}^{iA} &=&
                \frac{\delta}{\delta\psi_{iA}}.
\end{eqnarray}
How we should order the operators is the serious problem
in the quantum gravity.
While there is some discussion
to select some special ordering, we have no precise
answer to this problem a priori\cite{JS}.
So now we avoid this problem and simply
fix the operator ordering as in (\ref{eq:G})-(\ref{eq:Hi}).

We redefine two 1-form fields $\omega_{AB}$ and $\psi_{A}$ as
\begin{eqnarray}
                \omega_{AB} &=& \omega_{iAB}dx^{i}, \label{eq:omega} \\
                \psi_{A} &=& \psi_{iA}dx^{i},
\end{eqnarray}
where the index $i$ is the space index,\,$i=1,2,3$.
Then as we can easily see, the semi-classical solution for
the constraints (\ref{eq:G})-(\ref{eq:Hi}) are given by
\begin{equation}
                \Phi\kakko{\omega_{AB}, \psi_{A}} =
                \!\exp \kakko{-\frac{3}{2g^{2}}\int \left\{
                \!\omega_{AB}\wedge d \omega^{AB}
                \!+\frac{2}{3}\omega_{AC}\wedge\omega^{C}_{B}
                \wedge\omega^{AB}
                \!-\!\lambda g \psi^{A}\!\wedge
                \!D\psi_{A} \right\}  }. \label{eq:Phi}
\end{equation}
We call this the {\it{holomorphic wave function}} of the N=1 supergravity.
In the Einstein gravity, this type of the wave function is given in
Ref\cite{HK},
and has the form of the
exponential of the Chern-Simons functional. In the case of
the N=1 supergravity, the part of the Chern-Simons functional is replaced by
its supersymmetric extension; in fact, the functional
\begin{equation}
                W=\int \left\{
                \omega_{AB}\wedge d \omega^{AB}
                +\frac{2}{3}\omega_{AC}\wedge\omega^{C}_{B}
                \wedge\omega^{AB}
                -\lambda g \psi^{A}\wedge D\psi_{A} \right\}
\end{equation}
is invariant under the local supersymmetry transformation
\begin{eqnarray}
                \delta\omega_{AB} &=& -\lambda g \,\epsilon_{(A}
                \psi_{B)}, \\
                \delta\psi_{A} &=& D\epsilon_{A},
\end{eqnarray}
where the covariant derivative $D$ is that corresponding
to the connection (\ref{eq:omega}).

\section{The Robertson-Walker Universe}

In this section we consider the special case that
the space-time metric is given by the Robertson-Walker metric
and re-examine the discussion of the last section.

The Robertson-Walker metric is given by
\begin{equation}
                ds^{2} = -N^{2}dt^{2} \;\; + \;\;  \frac{1}{8} \;
                e^{2\alpha}\chi_{AB}\otimes\chi^{AB}, \label{eq:App4}
\end{equation}
where the 1-form $\chi_{AB}$ on
the 3-dim space (see appendix) satisfies the structure equation
\begin{equation}
                d\chi^{AB}=\chi^{A}_{C}
                \wedge \chi^{CB}, \label{eq:App5}
\end{equation}
and $N=N(t)$ and $\alpha=\alpha(t)$ depend only on the time.
In this metric the 3-dim space has the topology of $S^{3}$.

Now we suppose that the universe is homogeneous and isotropic, and
decompose the variables appearing in the theory
into the parts which depend only on the time and which
depend only on the space coordinates through $\chi_{AB}$:
\begin{eqnarray}
                \omega_{iAB} &=&i \omega \chi_{iAB},
                \nonumber \\
                \tilde{\pi}^{iAB} &=& - \frac{1}{24V} \sigma
                |\chi| \chi^{iAB}, \nonumber \\
                \psi_{iA} &=& \frac{i}{\sqrt{24V}}
                \chi_{iAB} \theta^{B}+\frac{i}{\sqrt{8V}}
                \Theta_{ABC} \chi^{BC}_{i},\nonumber \\
                \tilde{\pi}^{i}_{A} &=&
                \frac{1}{\sqrt{6V}} |\chi|
                \chi^{i}_{AB} \eta^{B}
                +\frac{1}{\sqrt{8V}} \Xi_{ABC}
                |\chi| \chi^{iBC} \label{eq:decomp},
\end{eqnarray}
where $\omega$, $\sigma$, $\theta_{A}$, $\Theta_{ABC}$,
$\eta_{A}$, and $\Xi_{ABC}$ are the variables depending only on the time.
We define the dual basis
$\chi^{i}_{AB}$ by $\chi_{iAB}\chi^{jAB}=8\delta^{j}_{i}$,
and $|\chi|$ is equal to $\det(\chi^{I}_{i})$,
where $\chi_{iAB}=\chi^{I}_{i}\tau_{IAB}$.
Here we normalize the
volume of $S^{3}$ as $V=\int d^{3}x |\chi|=\frac{\pi^{2}}{4}$.
The variable $\sigma$ is related to $\alpha$ as
\begin{equation}
                \sigma=12Ve^{2\alpha}.
\end{equation}
We assume that all the Lagrange multipliers $\omega_{0AB},
\psi_{0A}, M_{A}, N, $ and $N^{i}$ (where $N=\sqrt{q}\,\utilde{N}$ and
$M_{A}=\sqrt{q}\,\utilde{M}_{A}$. See appendix.)
depend only on the time.
Then the general solution (\ref{eq:Phi}) is rewritten as
\begin{eqnarray}
                &&\Phi\kakko{\omega,\theta_{A},\Theta_{ABC}}=
                e^{iS}, \label{eq:scw}  \\
                &&S = \frac{12}{g^{'2}}\kakko{
                \frac{\omega^{3}}{3}-\frac{i}{2}\omega^{2}
                +\frac{1}{4}\lambda' g'(\omega-i)
                \theta_{A}\theta^{A}
                -\frac{1}{4}\lambda' g'(\omega+2i)
                \Theta_{ABC}\Theta^{ABC}}.\nonumber \\ \label{eq:sca}
\end{eqnarray}

After integrating out the spatial coordinates, the Lagrangian
has the following form:
\begin{equation}
                L = \dot{\omega}\sigma
                +\dot{\theta}_{A}\eta^{A}
                +\dot{\Theta}_{ABC}\Xi^{ABC}-H. \label{eq:rl}
\end{equation}
The explicit form of the Hamiltonian $H$ will be given in the
next section after
some discussion.
The sets of the canonical variables are
$\ckakko{\omega, \sigma}, \ckakko{\theta_{A}, \eta^{A}}$,
and $\ckakko{\Theta_{ABC}, \Xi^{ABC}}$.
The Poisson brackets between the canonical variables are
\begin{eqnarray}
                \ckakko{\omega, \,\sigma} &=& 1, \label{eq:PB3}\\
                \ckakko{\theta_{A}, \,\eta^{B}} &=& -\delta^{B}_{A}, \\
                \ckakko{\Theta_{ABC}, \,\Xi^{DEF}} &=&
                -\delta^{D}_{(A}\delta^{E}_{B}\delta^{F}_{C)}. \label{eq:PB5}
\end{eqnarray}
In the quantization, we replace the Poisson brackets
(\ref{eq:PB3})-(\ref{eq:PB5}) with
the canonical (anti-)commutation relations;
\begin{eqnarray}
                \kakko{\omega, \,\sigma} &=& i, \\
                \ckakko{\theta_{A}, \,\eta^{B}} &=& -i\delta^{B}_{A}, \\
                \ckakko{\Theta_{ABC}, \,\Xi^{DEF}} &=&
                -i\delta^{D}_{(A}\delta^{E}_{B}\delta^{F}_{C)}.
\end{eqnarray}
We choose the representation in which the variables
$\omega$, $\theta_{A}$, and $\Theta_{ABC}$ are diagonalized:
\begin{eqnarray}
                \sigma &=& \frac{1}{i}\frac{\partial}{\partial\omega}, \\
                \eta^{A} &=& \frac{1}{i}\frac{\partial}{\partial\theta_{A}}, \\
                \Xi^{ABC} &=& \frac{1}{i}\frac{\partial}{\partial\Theta_{ABC}}.
\end{eqnarray}
Using these variables, the constraint ${\bf L}^{A}$ is rewritten as
\begin{equation}
                {\bf L}^{A}=-\frac{6}{\sqrt{6V}i}|\chi|
                (2\omega\eta^{A}-\lambda' g'\theta^{A}\sigma). \label{eq:L2}
\end{equation}
The function (\ref{eq:scw}) doesn't satisfy the constraint (\ref{eq:L2})
in general. To make (\ref{eq:scw}) have the left supersymmetry,
we must set $\Theta_{ABC}=0$ and $\Xi_{ABC}=0$. Then the function
\begin{eqnarray}
                &&\Phi\kakko{\omega,\theta_{A}}=
                e^{iS}, \label{eq:scw2}  \\
                &&S = \frac{12}{g^{'2}}\kakko{
                \frac{\omega^{3}}{3}-\frac{i}{2}\omega^{2}
                +\frac{1}{4}\lambda' g'(\omega-i)
                \theta_{A}\theta^{A}}, \label{eq:sca2}
\end{eqnarray}
satisfies the constraint (\ref{eq:L2}) and all the remaining constraints:
\begin{eqnarray}
                {\bf G}^{AB} &=& \frac{i}{6V}|\chi|
                (\theta^{A}\eta^{B}+\theta^{B}\eta^{A}), \\
                {\bf R}^{A} &=& \frac{1}{12V^{2}\sqrt{6V}}|\chi|^{2}
                \sigma^{2}\kakko{2(i-\omega)\theta^{A}
                +\frac{g'}{3\lambda'}\eta^{A}}, \\
                {\bf H}_{i} &=& -\frac{2}{3V}|\chi|\eta_{A}
                \kakko{2(i-\omega)\theta_{B}
                +\frac{g'}{3\lambda'}\eta_{B}}\chi^{AB}_{i}, \\
                {\bf H} &=& \frac{2}{3V^{2}}\sigma|\chi|^{2}
                \left[ \sigma \left( i\omega-\omega^{2}+\frac{g^{'2}}{12}
                \sigma-\frac{1}{4}\lambda' g'\theta_{A}\theta^{A} \right)
                +\eta_{A}\ckakko{2(i-\omega)\theta^{A}
                +\frac{g'}{3\lambda'}\eta^{A}} \right] .
\end{eqnarray}
Thus we obtain the semi-classical wave function
of the N=1 supergravity in the Robertson-Walker universe.

\section{The Classical Limit of the Quantum Universe}

Next we consider what classical universe
is involved in the semi-classical wave function (\ref{eq:scw2}). Note that
(\ref{eq:scw2}) has the form of the WKB wave function.
In the WKB approximation,
$S$ is the classical principal function of the dynamical system.
So all the informations about the classical universe involved in
(\ref{eq:scw2})
will be derived from (\ref{eq:sca2}). Since $S$ is the principal function,
it must satisfy the Hamilton-Jacobi equations:
\begin{eqnarray}
                \frac{\partial S}{\partial t}&+&
                H(\omega, \sigma, \theta_{A}, \eta^{A}) = 0, \label{eq:HJ1} \\
                \sigma &=& \frac{\partial S}{\partial \omega}
                =\frac{12}{g^{'2}} \left( \omega^{2}-i\omega
                +\frac{1}{4}\lambda' g' \theta_{A}\theta^{A} \right)
                , \label{eq:HJ2}  \\
                \eta^{A} &=& \frac{\partial S}{\partial \theta_{A}}
                =\frac{6\lambda'}{g'} (\omega-i)\theta^{A}. \label{eq:HJ3}
\end{eqnarray}
The Hamiltonian $H$ vanishes under (\ref{eq:HJ2}) and (\ref{eq:HJ3})
because all the constraints vanish under these two relations.
Therefore (\ref{eq:HJ1}) is derived from
(\ref{eq:HJ2})-(\ref{eq:HJ3}) and the fact that
$S$ doesn't depend on the time explicitly.
So we obtain the result that the classical universe contained in
(\ref{eq:scw2})
obeys the equations (\ref{eq:HJ2}) and (\ref{eq:HJ3}).

The constraints contained in the N=1 supergravity are all first class
constraints and the Hamiltonian $H$ is the linear combination of
these constraints. Therefore all the multipliers are left unfixed.
To fix these multipliers and introduce the time evolution,
we must fix some gauge. Here we take the following gauge:
\begin{eqnarray}
               && \omega_{0AB}=0, \qquad \psi_{0A}=0, \qquad
                M_{A}=0,  \nonumber \\
               && N=1, \qquad N^{i}=0.\label{eq:mul}
\end{eqnarray}
In this gauge the Hamiltonian is
\begin{equation}
                H =
                4\sqrt{12V}\sigma^{\frac{1}{2}}
                \left( i\omega-\omega^{2}+\frac{g^{'2}}{12}
                \sigma-\frac{1}{4}\lambda' g'\theta_{A}\theta^{A} \right)
                +4\sqrt{12V}\sigma^{-\frac{1}{2}}
                \eta_{A}\kakko{2(i-\omega)\theta^{A}+
                \frac{g'}{3\lambda'}\eta^{A}}. \label{eq:lag}
\end{equation}

Of course the classical solutions must obey
the Hamilton equations:
\begin{eqnarray}
                \frac{d\omega}{dt} &=&
                \ckakko{\omega, \,H}=\frac{g^{'2}}{3}
                \sqrt{12V}\sigma^{\frac{1}{2}}, \label{eq:He1} \\
                \frac{d\theta_{A}}{dt} &=&
                \ckakko{\theta_{A}, \,H}=8\sqrt{12V}\sigma^{-\frac{1}{2}}
                (\omega-i)\theta_{A}, \label{eq:He2} \\
                \frac{d\sigma}{dt} &=&
                \ckakko{\sigma, \,H}
                = 4\sqrt{12V}\kakko{\sigma^{\frac{1}{2}}
                (2\omega-i)+2\sigma^{-\frac{1}{2}}
                \eta_{A}\theta^{A}}, \label{eq:He3}  \\
                \frac{d\eta^{A}}{dt} &=&
                \ckakko{\eta^{A}, \,H}
                = 4\sqrt{12V}\sigma^{-\frac{1}{2}}
                \kakko{\frac{1}{2}\lambda' g'\sigma\theta^{A}
                +2(\omega-i)\eta^{A}}. \label{eq:He4}
\end{eqnarray}
where we use the relations (\ref{eq:HJ2})-(\ref{eq:HJ3}).
The equations (\ref{eq:He3}) and (\ref{eq:He4}) can be derived
from the equations (\ref{eq:HJ2}), (\ref{eq:HJ3}),
(\ref{eq:He1}) and (\ref{eq:He2}).

The reason why we should set $\Theta_{ABC}=\Xi_{ABC}=0$ can be understood
from another point of view. We can derive more general equations
corresponding to (\ref{eq:HJ2}), (\ref{eq:HJ3}),
(\ref{eq:He1}) and (\ref{eq:He2}) from (\ref{eq:gl}), (\ref{eq:H}),
(\ref{eq:Phi}) and (\ref{eq:mul}):
\begin{eqnarray}
                &&(D\psi_{A})_{ij}+\frac{g}{3\lambda}
                \epsilon_{ijk}\tilde{\pi}^{k}_{A}=0, \\
                &&R_{ABij}-\frac{g^{2}}{3}\epsilon_{ijk}
                \tilde{\pi}^{k}_{AB}+\lambda g
                \psi_{i(A}\psi_{B)j}=0, \label{eq:ghj} \\
                &&\frac{\partial\omega_{iAB}}{\partial t}=
                \frac{ig^{2}}{6\sqrt{q}}\epsilon_{ijk}
                \tilde{\pi}^{j}_{AC}\tilde{\pi}^{kC}_{B}, \\
                &&\frac{\partial\psi_{iA}}{\partial t}=\frac{i}{\sqrt{q}}
                \tilde{\pi}^{j}_{AB}(D\psi^{B})_{ij}.
\end{eqnarray}
All the remaining Euler-Lagrange equations are derived from the above
equations. In the case of the Robertson-Walker universe, (\ref{eq:ghj})
gives the following three equations by the spinorial decomposition.
\begin{eqnarray}
                &&\sqrt{3}\Theta_{E(AB}\Theta_{CD)}^{E}
                +\Theta_{(ABC}\theta_{D)}=0, \\
                &&\Theta_{ABC}\theta^{C}=0, \\
                &&\omega^{2}-i\omega-\frac{g^{'2}}{12}\sigma
                +\frac{1}{4}\lambda' g'\theta_{A}\theta^{A}
                -\frac{1}{4}\lambda' g'\Theta_{ABC}\Theta^{ABC}=0.
\end{eqnarray}
The first and the second algebraic constraints
can be derived from neither the Euler-Lagrange
equations of (\ref{eq:rl}) nor the Hamilton-Jacobi
equations of (\ref{eq:sca}).
So when we consider the mini-superspace of the Robertson-Walker universe,
these extra constraints must not appear and we must
set $\Theta_{ABC}=\Xi_{ABC}=0$.
Of course $\Theta_{ABC}$ and $\Xi_{ABC}$ may be non-zero in the general case.

In our gauge the line element is
\begin{equation}
                12Vds^{2} = -d\tau^{2}+\frac{\sigma}{8}d^{2}\Omega,
                \label{eq:le}
\end{equation}
where $d^{2}\Omega \;\;$ is $\;\; \chi_{AB}\otimes\chi^{AB} \;\;$
and $\;\; d\tau=\sqrt{12V}dt$.
$\sigma$ must be real when (\ref{eq:le}) represents the Lorentzian or
the Euclidean universe. Then there exist four cases:
\begin{equation}
                  \begin{array}{cc}
             \left.  \begin{array}{lll}
                        \mbox{case 1} &  \tau:\mbox{real,} & \; \sigma<0 \\
                        \mbox{case 2} & \tau:\mbox{imaginary,} & \; \sigma>0
                     \end{array} \right\} & \mbox{Euclidean,} \\
             \left.  \begin{array}{lll}
                        \mbox{case 3} & \tau:\mbox{real,} & \; \sigma>0 \\
                        \mbox{case 4} & \tau:\mbox{imaginary,}  & \; \sigma<0
                     \end{array} \right\} & \mbox{Lorentzian.}
                   \end{array}
\end{equation}
We examine these four cases respectively.
Using (\ref{eq:HJ2}) and (\ref{eq:HJ3}), the equation (\ref{eq:He3})
is rewritten as
\begin{equation}
                \frac{d\sigma}{d\tau}=4\kakko{\sigma^{\frac{1}{2}}
                (6\omega-5i)-\frac{48}{g^{'2}} \sigma^{-\frac{1}{2}}
                \omega(\omega-i)^{2}}. \label{eq:mos}
\end{equation}
This equation and (\ref{eq:He1}) contain only the bosonic parameters.
When we search for the classical solutions,
first we solve these two equations and obtain $\omega$ and $\sigma$.
Then we use the results to solve (\ref{eq:He2}) and (\ref{eq:HJ3}).

{\it{Case 1}}: We set $\sigma=-r$. From (\ref{eq:He1}), $\omega$
has the form of
\begin{equation}
                \omega=c+if,
\end{equation}
where $c$ is the constant to be determined and
$f$ is the function depending only on $\tau$.
Then (\ref{eq:He1}) and (\ref{eq:mos}) give the following equations:
\begin{eqnarray}
                &&\frac{df}{d\tau}=\frac{g^{'2}}{3}r^{\frac{1}{2}}, \\
                &&\frac{dr}{d\tau}  =
                4\kakko{r^{\frac{1}{2}}(6f-5)+\frac{48}{g^{'2}}
                r^{-\frac{1}{2}}\ckakko{2c^{2}(f-1)+c^{2}f
                -f(f-1)^{2}}}, \\
                &&c\kakko{r+\frac{8}{g^{'2}}\ckakko{
                c^{2}-(f-1)(3f-1)}}=0.
\end{eqnarray}
We must set $c=0$
by the requirement of the consistency among the above equations.
There exists only the trivial solution when $c$ takes the non-zero value.
The equations satisfied by the classical solutions in this case are:
\begin{eqnarray}
                \frac{df}{d\tau} &=&
                \frac{g^{'2}}{3}r^{\frac{1}{2}}, \\
                \frac{dr}{d\tau} &=&
                4\kakko{r^{\frac{1}{2}}(6f-5)-\frac{48}{g^{'2}}
                r^{-\frac{1}{2}}f(f-1)^{2}}, \\
                \frac{d\theta_{A}}{d\tau} &=&
                8r^{-\frac{1}{2}}(f-1)\theta_{A}, \\
                \eta^{A}&=&\frac{6i\lambda'}{g'}(f-1)\theta^{A}.
\end{eqnarray}
It seems very difficult to solve these equations analytically.
We have been able to find only two analytic solutions.
Let us consider the case in which $f^{2}-f-\frac{g'^{2}}{12} r=0$.
This condition means that the gravity can be solved analytically by itself and
the gravitino exists in the background of the gravity.
Then the equations to be solved are
\begin{eqnarray}
                \frac{df}{d\tau} &=&
                \frac{g^{'2}}{3}r^{\frac{1}{2}}, \label{eq:c11} \\
                r&=&\frac{12}{g^{'2}}
                \kakko{\left(f-\frac{1}{2} \right)^{2}-\frac{1}{4}}, \\
                \frac{d\theta_{A}}{d\tau} &=&
                8r^{-\frac{1}{2}}(f-1)\theta_{A}, \\
                \eta^{A} &=&
                \frac{6i\lambda'}{g'}(f-1)\theta^{A}.
\end{eqnarray}
There exist two cases:
\begin{eqnarray}
                (a):\,\,\,  f&=& \frac{1}{2}(1+\cosh\xi), \nonumber \\
                (b):\,\,\,  f&=& \frac{1}{2}(1-\cosh\xi),
\end{eqnarray}
where $\xi$ is the function of $\tau$ to be determined.
In either case, $r$ is given by
\begin{equation}
                r=\frac{3}{g^{'2}}\sinh^{2}\xi.
\end{equation}
First we consider case $(a)$.
{}From (\ref{eq:c11}), we have
\begin{equation}
                \xi=\pm\frac{2}{\sqrt{3}}|g'|\tau+const.
\end{equation}
We set $\xi=\frac{2}{\sqrt{3}}|g'|\tau$. Then the inequality $\xi\geq0$
must be satisfied. The solution is
\begin{eqnarray}
                12Vds^{2} &=& -\frac{3}{4g^{'2}} \left( d\xi^{2}
                +\frac{1}{2}\sinh^{2}\xi \,\,d^{2}\Omega \right),\\
                f&=& \frac{1}{2}(1+\cosh\xi),\\
                \theta_{A} &=& \cosh^{4}\frac{\xi}{2}\,\,\theta_{0A}, \\
                \eta^{A} &=& \frac{6i\lambda'}{g'}
                \sinh^{2}\frac{\xi}{2}\,\,\cosh^{4}\frac{\xi}{2}
                \,\,\theta_{0}^{A},
\end{eqnarray}
where the grassmannian constant $\theta_{0A}$ satisfies
$\theta_{0A}\theta_{0}^{A}=0$.
This solution covers the region $\tau\geq0$ and has the topology of
the hyperbolic universe $H^{4}$.

The case $(b)$ gives another analytic solution
which covers the region $\tau\leq0$. By the similar discussion, we obtain
\begin{eqnarray}
                \xi &=& \frac{2}{\sqrt{3}}|g'|\tau, \,\,\,\,\,\xi\leq0, \\
                12Vds^{2} &=& -\frac{3}{4g^{'2}}( d\xi^{2}
                +\frac{1}{2}\sinh^{2}\xi \,\,d^{2}\Omega),\\
                f&=& \frac{1}{2}(1-\cosh\xi),\\
                \theta _{A} &=& \sinh^{4}\frac{\xi}{2}
                \,\,\theta_{0A}, \\
                \eta^{A} &=& -\frac{6i\lambda'}{g'}
                \cosh^{2}\,\,\frac{\xi}{2}
                \,\,\sinh^{4}\frac{\xi}{2}\,\,\theta_{0}^{A}.
\end{eqnarray}
where $\theta_{0A}\theta_{0}^{A}=0$.

{\it{Case 2}}: We set $\tau=i\eta$, where $\eta$ is a real parameter.
$\omega$ is written as $\omega=c+if(\eta)$. Then we have
\begin{eqnarray}
               &&\frac{df}{d\eta} =
               \frac{g^{'2}}{3}\sigma^{\frac{1}{2}}, \\
               &&\frac{d\sigma}{d\eta} =
               4\kakko{\sigma^{\frac{1}{2}}(5-6f)
               +\frac{48}{g^{'2}}\sigma^{-\frac{1}{2}}
               \ckakko{(3f-2)c^{2}-f(f-1)^{2}}}, \\
               &&c\kakko{\sigma-\frac{8}{g^{'2}}
               \ckakko{c^{2}-(f-1)(3f-1)}}=0.
\end{eqnarray}
By the requirement of the consistency among these equations,
we must set $c=0$. Otherwise we obtain only the trivial solution.
The equations to be solved are:
\begin{eqnarray}
               &&\frac{df}{d\eta} =
               \frac{g^{'2}}{3}\sigma^{\frac{1}{2}}, \\
               &&\frac{d\sigma}{d\eta} =
               4\kakko{\sigma^{\frac{1}{2}}(5-6f)
               -\frac{48}{g^{'2}}\sigma^{-\frac{1}{2}}
               f(f-1)^{2}}, \\
               &&\frac{d\theta_{A}}{d\eta}=
               -8\sigma^{-\frac{1}{2}}(f-1)\theta_{A}, \\
               &&\eta^{A} = \frac{6i\lambda'}{g'}(f-1)\theta^{A}.
\end{eqnarray}
As is in the case1, we can obtain the analytic solution when the condition
$f^{2}-f+\frac{g'^{2}}{12}\sigma=0$ is satisfied:
\begin{eqnarray}
                \xi &=& \frac{2|g'|}{\sqrt{3}}\eta, \,\,\,\,\,
                0\leq\xi\leq\pi,  \\
                12Vds^{2} &=& \frac{3}{4g^{'2}}\left( d\xi^{2}
                +\frac{1}{2}\sin^{2}\xi \,\,d^{2}\Omega \right), \\
                f &=& \frac{1}{2}(1-\cos\xi), \\
                \theta_{A} &=& \sin^{4}\frac{\xi}{2}\,\,\theta_{0A}, \\
                \eta^{A} &=& -\frac{6i\lambda'}{g'}
                \cos^{2}\frac{\xi}{2}\,\,
                \sin^{4}\frac{\xi}{2}\,\,\theta_{0}^{A},
\end{eqnarray}
where $\theta_{0A}\theta_{0}^{A}=0$, and the universe has the topology
of the sphere $S^{4}$.

{\it{Case 3}}: In this case we set $\omega=f(\tau)+ic$.
Then we have
\begin{eqnarray}
                \frac{df}{d\tau} &=& \frac{g^{'2}}{3}\sigma^{\frac{1}{2}}, \\
                \frac{d\sigma}{d\tau} &=&
                24f\kakko{\sigma^{\frac{1}{2}}
                -\frac{8}{g^{'2}}\sigma^{-\frac{1}{2}}\ckakko{
                f^{2}-(c-1)(3c-1)}}, \\
                \sigma &=& \frac{48}{g^{'2}}\frac{1}{6c-5}
                \kakko{f^{2}(3c-2)-c(c-1)^{2}}.\label{eq:c31}
\end{eqnarray}
By the consistency, we obtain $c=\frac{1}{2}$. The equation (\ref{eq:c31})
gives
\begin{equation}
                \sigma=\frac{12}{g^{'2}}\left( f^{2}+\frac{1}{4} \right).
\end{equation}
Because of this relation, there exists only one solution in this case:
\begin{eqnarray}
                \xi &=& \frac{2|g'|}{\sqrt{3}}\tau, \\
                12Vds^{2} &=& \frac{3}{4g^{'2}}
                \kakko{-d\xi^{2}+\frac{1}{2}\cosh^{2}\xi\, d^{2}\Omega}, \\
                f &=& \frac{1}{2}\sinh\xi, \\
                \theta_{A} &=&
                \cosh^{2}\xi\, \exp\,
                \kakko{-2i\,\arctan \,\sinh \xi}\theta_{0A}, \\
                \eta^{A} &=&
                \frac{3\lambda'}{g'}(\sinh\xi-i)
                \cosh^{2}\xi \,\exp\,
                \kakko{-2i\,\arctan \, \sinh \xi}\theta_{0}^{A}.
\end{eqnarray}
This universe has the topology of the de Sitter universe $dS^{4}$.

{\it{Case4}}:  We set $\tau=i\eta$, $\sigma=-r\,\,(r\geq0)$,
and $\omega=f(\eta)+ic$. Then we have
\begin{eqnarray}
                \frac{df}{d\eta} &=& -\frac{g^{'2}}{3}r^{\frac{1}{2}}, \\
                \frac{dr}{d\eta} &=&
                4f\kakko{6r^{\frac{1}{2}}+\frac{48}{g^{'2}}
                r^{-\frac{1}{2}}\ckakko{f^{2}-(c-1)(3c-1)}}, \\
                r &=& -\frac{48}{g^{'2}}\frac{1}{6c-5}
                \kakko{(3c-2)f^{2}-c(c-1)^{2}}.\label{eq:c41}
\end{eqnarray}
By the consistency we get $c=\frac{1}{2}$. Using this value (\ref{eq:c41})
is rewritten as
\begin{equation}
                r=-\frac{12}{g^{'2}}\left(f^{2}+\frac{1}{4} \right)<0,
\end{equation}
which is inconsistent with $r\geq0$. Therefore there exists no solution
in case 4.


Now we examine the differential equations obtained above.
If we suitably change the normalization of
$\sigma$, $f$, $\theta_{A}$, and $\eta_{A}$,
we can absorb $g'$ and $\lambda'$. So we calculate under the condition
that $g'=1$ and $\lambda'=1$.
Since $f$ is the monotone increasing function of $\tau$ (or $\eta$),
we may consider $f$ as the time parameter.
Since $\theta_A$ and $\eta_A$ can be written as
\begin{eqnarray}
&& \theta_A = F_{\theta} \theta_{0A}, \\
&& \eta_A = F_{\eta} \eta_{0A},
\end{eqnarray}
where $F_{\theta}$ and $F_{\eta}$ are the c-number functions
depending on $f$, and $\theta_{0A}$ and $\eta_{0A}$
are the grassmannian constants,
we calculate and plot the bosonic parameters $F_{\theta}$ and $F_{\eta}$
under some suitable normalizations.

In the case 3, we have the analytic
solution, and in the case 4, there exists no solution.
So we pay attention mainly to the case 1 and the case 2.
In these cases it
seems difficult to solve the differential equations
analytically in general,
though we can find the analytic solutions
in some special conditions.
Therefore we resort to the numerical calculation to get the solutions,
and study the nature of these solutions.

In the case 1 and case 2,
using the Runge-Kutta method, we calculated the solution curves numerically.
Since we set $c=1, \hbar=1,$ and $G=1$, all quantities appearing in
graphs are dimensionless.
The fig.\ref{fig:case1} is the graph of $r$ (case 1). Corresponding to
some of these solutions (labeled by A,B,C,D,E and F), the graphs of
$F_{\theta}$ and $F_{\eta}$ are given by the fig.\ref{fig:case1t} and
the fig.\ref{fig:case1e}.
The fig.\ref{fig:case2} is the graph of $\sigma$ (case 1), and
the graphs of
$F_{\theta}$ and $F_{\eta}$ are given in the
fig.\ref{fig:case2t},\ref{fig:case2e}
(for G,H,I,and J).
The solutions B,F and J are the analytic solutions.
In the case 3 we draw the graph of the analytic solution
(fig.\ref{fig:case3}-fig.\ref{fig:case3t}).

In these results we can see some new properties
which come from the existence of the gravitino.
In the case 1 we have some solutions that $r \rightarrow \infty$
as $f \rightarrow \pm \infty$ (for example B,C,..,F).
We also have some solutions
which seem to correspond to the compact universes (A and G),
though the pure gravity solution in the case 1
has the topology of the hyperbolic non-compact universe.
{}From fig.\ref{fig:case1}-fig.\ref{fig:case1e} we can see that
the very rapid increase of $F_{\theta}$ and $F_{\eta}$ may be the
causes of these compactification of the universes.

There are the universes which seem to have the singularities
in the first and/or the end.
The scalar curvature of the Robertson-Walker universe
whose metric is given by (\ref{eq:App4}) with
the lapse function $N=1$
is calculated as
\begin{equation}
R= 6 \left( 4 e^{-2\alpha} +\frac{d^2 \alpha}{dt^2}
            + 2\left( \frac{d \alpha}{dt} \right)^2 \right).
\end{equation}
Using the relations $12V e^{2 \alpha} = \sigma $, $V=\pi^2/4$, and
$d\tau = \sqrt{3\pi^2} dt$, we have
\begin{equation}
R= 9\pi^2 \fra{8+\fra{d^2\sigma}{d\tau^2}}{\sigma}.
\end{equation}
The graphs of the scalar curvature of the solutions A,B,C,D,E and F
are given in the fig.\ref{fig:case1r},
and that of G,H,I and K in the fig.\ref{fig:case2r}.

How we should take the direction of the physical time is the serious problem.
In the analytic solutions in the case 1,
we have two candidates of the time ; $\xi$ and $\tau$.
We can take $\xi$ as the time
because the universe expands with the increase of $\xi$
(Note that the cosmological constant of the N=1 supergravity is positive.).
But we have the solutions which diverge in the limit $f(or\, \tau)\rightarrow
\pm \infty$ in the case 1,
and can't decide the direction in the case 2 because all the universes
in this case are compact.
So we can't decide the direction of the time naively.


\section{Discussion}

In this paper we considered the non-perturbative canonical quantization
of the N=1 supergravity with the cosmological terms by
the Ashtekar formalism. We obtained the holomorphic wave function
of the universe which is given by the exponential
of the N=1 supersymmetric extension of the Chern-Simons functional.
We applied this wave function to the Robertson-Walker metric
and found that there exist several types of
the universe which contain the gravitinos.
Furthermore we obtained four exact classical solutions.
The gravitinos in the numerical and the analytic solutions can't be deleted
by the supersymmetry transformation in general.

The N=1 supergravity in the Ashtekar formalism is the complex theory
and we must consider the reality conditions (see \cite{J})
to take out the real solutions. In this paper we have considered
the only one condition that the dreibein fields must be real.
When we consider the Lorentzian (Euclidean) universe,
the action must be real (pure imaginary).
The reality condition about the $SL(2,C)$ connection in the
Lorentzian universe is the
torsion condition, which is derived from the reality of the
action:
\begin{equation}
                {\cal{D}}\,e_{AA'}=\psi_{A}\wedge\psi_{A'},
\end{equation}
where ${\cal{D}}$ is the covariant derivative which acts on
both left and right spinor indices. But as can be seen easily,
the action is real for the classical solution of the case 3
and pure imaginary for that of the case 2 or 3. Therefore
we may think that the reality condition on the connection is satisfied.
The reality condition on the gravitino is the Majonara condition.
The classical solutions in this paper don't satisfy this condition in general.
To obtain the real solution, we must transform the solution by the
transformations corresponding to the symmetries
(local Lorentz, left and right supersymmetries,
3-dim diffeomorphism, and time-reparametrization)
in the theory. Since it seems very difficult to determine
the parameters of the transformations explicitly and ascertain whether
there exist non-trivial solutions,
we leave the settlement of the problem about the reality conditions
and will consider it on another occasion.

In this paper we used the self-dual representation. We know another
approach to the quantum gravity, which is called the loop space representation
\cite{JS}. Whether there exists the corresponding representation
in the N=1 supergravity is the interesting and challenging problem.
In the Einstein gravity, the physical states in the loop space representation
are related to the invariants of knots.
It seems natural that we expect some invariants corresponding to the
physical states of the N=1 supergravity.

Recently the Ashtekar formalism of the N=2 supergravity is derived \cite{KS}.
We can obtain the holomorphic wave function of the N=2 supergravity as well as
that of the N=1 supergravity.
 The study in this case is now in progress.

\section*{Acknowledgement}
The authors thank Professor T. Eguchi and Dr. H. Kunitomo
for helpful discussions and useful comments on the manuscript.

\section*{Appendix}

Here we give the notations and formulas used in this paper.
Our space-time signature is $(-,+,+,+)$. We represent the 4-dim
space-time indices and the local Lorentz indices by $\mu, \nu, \rho,\cdots,$
and $a, b, c,\cdots,$ and the 3-dim space indices and the flat space indices by
$i, j, k,\cdots,$ and $I, J, K,\cdots,$ respectively.
we take the basis of
$SL(2,C)$ and $SU(2)$ spinors as
\begin{equation}
                \begin{array}{cc}
                \sigma^{0}_{AA'}=\fra{1}{\sqrt{2}}
                \left(  \begin{array}{cc}
                1 & 0 \\
                0 & 1 \end{array}
                \right),&
                \sigma^{1}_{AA'}=\fra{1}{\sqrt{2}}
                \left(  \begin{array}{cc}
                0 & 1 \\
                1 & 0 \end{array}
                \right),\\
                \sigma^{2}_{AA'}=\fra{1}{\sqrt{2}}
                \left(  \begin{array}{cc}
                0 & i \\
                -i & 0 \end{array}
                \right),&
                \sigma^{3}_{AA'}=\fra{1}{\sqrt{2}}
                \left(  \begin{array}{cc}
                1 & 0 \\
                0 & -1 \end{array}
                \right),
                \end{array}
\end{equation}
and
\begin{equation}
                \begin{array}{ccc}
                \tau^{1}_{AB}=
                2i\left(  \begin{array}{cc}
                1 & 0 \\
                0 & -1 \end{array}
                \right), &
                \tau^{2}_{AB}=
                2i\left(  \begin{array}{cc}
                i & 0 \\
                0 & i \end{array}
                \right), &
                \tau^{3}_{AB}=
                2i\left(  \begin{array}{cc}
                0 & -1 \\
                -1 & 0 \end{array}
                \right),
                \end{array}
\end{equation}
respectively. By these basis, we can transform the $SO(3,1)$ vector $v_{a}$
into $SL(2,C)$ spinor as $v_{AA'}:=v_{a}\sigma^{a}_{AA'}$, and
the $SO(3)$ vector $u_{I}$ into $SU(2)$ spinor as $u_{AB}:=u_{I}\tau^{I}_{AB}$.
We define the anti-symmetric spinors by
\begin{equation}
                \epsilon_{AB}=\epsilon^{AB}
                =\epsilon_{A'B'}=\epsilon^{A'B'}
                =\left(  \begin{array}{cc}
                0 & 1 \\
                -1 & 0 \end{array} \right).
\end{equation}
The spinor indices can be raised and lowered according to the conventions
\begin{equation}
                \lambda^{A}=\epsilon^{AB}\lambda_{B},
                \,\,\,\lambda_{A}=\lambda^{B}\epsilon_{BA}.
\end{equation}

Taking an adequate gauge, we fix the form of the vierbein field as follows:
\begin{equation}
                {e_{\mu}}^{a}=
                \left(
                \begin{array}{cccc}
                N && N^{j}e_{j}^{I} & \\
                  &&& \\
                0 && e_{i}^{I} & \\
                  &&&
                \end{array}
                \right),
\end{equation}
where $N$ and $N^{i}$ are the lapse function and the shift vector,
respectively.
Defining the 3-dim space metric $q_{ij}$ by $q_{ij}=e^{I}_{i}e_{Ij}$,
the line-element of space-time is given by
\begin{equation}
                ds^{2}=-N^{2}dt^{2}+q_{ij}
                (N^{i}dt+dx^{i})(N^{j}dt+dx^{j}).
\end{equation}
Introducing the dual basis $e^{i}_{I}$ by $e^{i}_{I}e_{j}^{I}=\delta^{i}_{j}$,
we have
\begin{equation}
                \tilde{\pi}^{i}_{AB}=
                -\frac{1}{2}\sqrt{q}e^{i}_{I}\tau^{I}_{AB}
\end{equation}
by the straightforward calculation, where $q=\det q_{ij}$.
We also have
\begin{eqnarray}
                \Sigma^{AB}_{0i}&=&
                \frac{1}{2}Nie_{iI}\tau^{I AB}+\frac{1}{2}\epsilon_{ijk}
                N^{j}\sqrt{q}e^{k}_{I}\tau^{I AB} \\
                &=&-\utilde{N}i\tilde{\pi}^{AB}_{i}
                -\epsilon_{ijk}N^{j}\tilde{\pi}^{kAB},
\end{eqnarray}
where $\utilde{N}=N/\sqrt{q}$.
By the formula $\tilde{\pi}_{iAB}=-\frac{1}{2}\epsilon_{ijk}
\tilde{\pi}^{j}_{AC}\tilde{\pi}^{kC}_{B}$, finally we obtain
\begin{equation}
                \Sigma^{AB}_{0i}=
                -\frac{1}{2}\epsilon_{ijk}
                \kakko{-i\;\utilde{N}\tilde{\pi}^{jA}_{C}
                \tilde{\pi}^{kCB}+2N^{j}\tilde{\pi}^{kAB}}.
\end{equation}
We can ascertain that (\ref{eq:App3}) is the solution of (\ref{eq:App1})
under (\ref{eq:App2}). By counting of
the degrees of freedom, (\ref{eq:App3}) is just the general solution.

Now we list some formulas in the case of the Robertson-Walker metric
(\ref{eq:App4}) and (\ref{eq:App5}).
The space-time metric is given by
\begin{equation}
                g_{\mu\nu}= \left(
                \begin{array}{cccc}
                -N^{2} && 0 &\\
                &&& \\
                0 && \frac{1}{8}e^{2\alpha}\chi_{iAB}\chi^{AB}_{j} & \\
                &&&
                \end{array}  \right)
                =\left(
                \begin{array}{cccc}
                -N^{2} && 0 &\\
                &&& \\
                0 && e^{2\alpha}\chi^{I}_{i}\chi_{Ij} & \\
                &&&
                \end{array}  \right),
\end{equation}
and the vierbein is given by
\begin{equation}
                {e_{\mu}}^{a}=
                \left(
                \begin{array}{cccc}
                N && 0 & \\
                  &&& \\
                0 && e^{\alpha}\chi^{I}_{i}& \\
                  &&&
                \end{array}
                \right).
\end{equation}
Moreover we obtain
\begin{eqnarray}
                e:=\det e_{\mu}^{a}=Ne^{3\alpha}|\chi|, \\
                \sqrt{q}=e^{3\alpha}|\chi|, \\
                \tilde{\pi}^{i}_{AB}=-\frac{1}{2}|\chi|
                e^{2\alpha}\chi^{i}_{AB}.
\end{eqnarray}
In this paper we use the Maurer-Cartan form of $SU(2)$ as $\chi_{i}^{AB}$;
\begin{eqnarray}
                {U^{A}}_{B}&=&
                \left( \begin{array}{cc}
                       e^{-\frac{i}{2}(\alpha+\gamma)}\cos\frac{\beta}{2} &
                       -e^{-\frac{i}{2}(\alpha-\gamma)}\sin\frac{\beta}{2} \\
                       \\
                       e^{\frac{i}{2}(\alpha-\gamma)}\sin\frac{\beta}{2} &
                       e^{\frac{i}{2}(\alpha+\gamma)}\cos\frac{\beta}{2}
                \end{array}    \right), \\
                \nonumber \\
                {\chi^{A}}_{B}&=&-{{U^{-1}}^{A}}_{C} \,{dU^{C}}_{B},
\end{eqnarray}
where $\alpha$, $\beta$, and $\gamma$ are the Euler angles.
The explicit form of ${\chi_{i}}^{I}$ is
\begin{equation}
                {\chi_{i}}^{I}=
                \left(        \begin{array}{ccc}
                     \frac{1}{4}\sin\beta\cos\gamma &
                     -\frac{1}{4}\sin\beta\sin\gamma &
                     -\frac{1}{4}\cos\beta \\
                     && \\
                     -\frac{1}{4}\sin\gamma &
                     -\frac{1}{4}\cos\gamma & 0 \\
                     && \\
                     0 & 0 & -\frac{1}{4}
                     \end{array}   \right),
\end{equation}
and then we have
\begin{eqnarray}
                &&|\chi| = \frac{1}{64}\sin\beta, \\
                &&\partial_{i}(\,|\chi|\chi_{AB}^{i}\,)=0.
\end{eqnarray}

\newpage

\begin{figure}
\caption{The graphs of $r$ in the case 1.}
\label{fig:case1}
\end{figure}

\begin{figure}
\caption{The graphs of the real part of $F_{\theta}$ in the case 1}
\label{fig:case1t}
\end{figure}

\begin{figure}
\caption{The graphs of the imaginary part of $F_{\eta}$ in the case 1}
\label{fig:case1e}
\end{figure}

\begin{figure}
\caption{The graphs of $\sigma$ in the case 2}
\label{fig:case2}
\end{figure}

\begin{figure}
\caption{The graphs of the real part of $F_{\theta}$ in the case 2}
\label{fig:case2t}
\end{figure}

\begin{figure}
\caption{The graphs of the imaginary part of $F_{\eta}$ in the case 2}
\label{fig:case2e}
\end{figure}

\begin{figure}
\caption{The graphs of $\sigma$ in the case 3}
\label{fig:case3}
\end{figure}

\begin{figure}
\caption{The graphs of $F_{\theta}$ in the case 3}
\label{fig:case3t}
\end{figure}

\begin{figure}
\caption{The graphs of $F_{\eta}$ in the case 3}
\label{fig:case3e}
\end{figure}

\begin{figure}
\caption{The graphs of the scalar curvature $R$ in the case 1}
\label{fig:case1r}
\end{figure}

\begin{figure}
\caption{The graphs of the scalar curvature $R$ in the case 2}
\label{fig:case2r}
\end{figure}

\end{document}